\documentclass[a4paper,aps,superscriptaddress,nofootinbib,floatfix,epfs,12pt]{article}
\usepackage[margin=2.75cm]{geometry}
\usepackage{amsfonts,amscd,amsmath,amssymb,graphicx,graphbox,color,float,relsize,authblk}

\def\be{\begin{equation}}
\def\ee{\end{equation}}
\def\beq{\begin{eqnarray}}
\def\eeq{\end{eqnarray}}

\def\parline{\,\partial\kern -0.55em /\,\,}

\def\half{{\frac{1}{2}}}

\def\AA{{\cal A}}

\def\CC{{\cal C}}
\def\DD{{\cal D}}

\def\MM{{\cal M}}

\def\RR{{\cal R}}

\def\ZZ{{\cal Z}}

\def\sm(A)dS{{\scriptscriptstyle (A)dS }}

\def\zb{\bar{z}}

\def\irm{{\rm i}}

\begin{document}

\title{Covariant calculation of the partition function \\ of the two-dimensional sigma model \\ on compact two-surfaces}
\author{O.D. Andreev, R.R. Metsaev, and A.A. Tseytlin}
\affil{Department of Theoretical Physics, P.N. Lebedev Physical Institute, Leninski prospect 53, Moscow  117924, USSR}
\date{\small{(submitted 17 July 1989)}}
\maketitle
\begin{abstract} 
Motivated by string theory connection, a  covariant procedure for perturbative calculation of the partition function $Z$ of the two-dimensional generalized $\sigma$-model is considered. The importance of a consistent  regularization of the measure in the path  integral is emphasized. The partition function $Z$ is computed  for a number of specific 2-manifolds: sphere, disk and torus. 
\end{abstract}

\

\

\begin{flushleft}{Published in:  \ \ Yad.Fiz. 51 (1990) 564-576\ \  [Sov.J.Nucl.Phys. 51 (1990) 359-366] } \end{flushleft}

\def \RR {{\rm R}}\def \CC {{\rm C}}

\newpage

\tableofcontents
\setcounter{footnote}{0}
\setcounter{section}{0}

\section{Introduction}
A promising approach to string theory is the so-called $\sigma$-model approach. It may help 
 elucidate the structure and first principles of string theory  (see, e.g., Refs.\cite{1,2}). 

A central role in the $\sigma$-model approach is played by the partition function $Z$ of the generalized 
two-dimensional $\sigma$-model. $Z$ is closely related to the generation functional for the string S-matrix and to the effective action of the string theory \cite{2}.

The string partition function differs from the usual $\sigma$-model partition function by a factor of the volume of the M\"obius group. In the theory of closed strings a possible implementation 
 of the operation of division by the M\"obius group volume is by   taking the derivative with respect to the  log of the UV cutoff $\partial\over \partial\ln\varepsilon$ of the regularized partition function $Z_R$. The reason for this  is the presence of a logarithmic  divergence  \cite{3} in the regularized volume $\Omega_R$ of the M\'obius group \cite{4}. 
 
 In the theory of open strings the procedure of  ``division" by the M\"obius volume reduces to a renormalization of  power divergencies as the regularized volume of the M\"obius group $SL(2,\RR)$ contains only power divergences  \cite{5,3}. The remaining logarithmic two-dimensional UV divergences can be interpreted as being due to the massless poles in the scattering amplitudes. As a result, the renormalized string partition function coincides with the effective action $S$ for the massless modes of the open string.

We shall perform the calculation of the partition function of the two-dimensional $\sigma$-model on compact surfaces emphasizing  the role of the measure in the functional integral in the procedure of calculating the covariant expression for $Z$. In Sec.2 we consider three possible ways of determining the regularized measure that lead to a covariant answer. In Sec.3 we give examples of the calculation of the leading terms in $Z$ for some specific cases of 2-manifolds: the sphere, disk (hemisphere), and the torus. Taking into account the procedure for dividing by the M\"obius volume, we obtain an alternative to the S-matrix method of \cite{6} for calculating the string effective action.
In Sec.4 we consider a generalization of this  approach to the  supersymmetric case. 

Let us  make a comment on the interpretation of   infinities that are present in $Z$. In addition to the already mentioned M\"obius and other two-dimensional UV infinities, in the case of $2$-surfaces of higher  genera there exist the so-called modular infinities corresponding to degeneration of the Riemann surfaces \cite{9}.\footnote{In the framework of the $\sigma$-model approach,  in the case of surfaces of higher genera it is necessary to use  the Schottky  \cite{7} or  the branch-point  type \cite{8}  parameterization for the  moduli 
space  in which the on-shell scattering amplitudes have  formal $SL(2,\CC)$ invariance.} The ``modular' correction to the $\beta$-functions corresponds to the infinities associated with the degeneration of trivial cycles \cite{10}. The partition function $Z$ should be renormalizable with respect to all infinities (modular and local), i.e. it  should be finite after the renormalization corresponding to the complete $\beta$-function \cite{11}.

\def \AA  {{\ZZ_0}}
\def \na {\nabla}
\def \D   {{\mathcal D}}
\def \rd {{\rm D}}

\section{Calculation of  partition function of  $\sigma$-model on compact 2-surfaces}
\renewcommand{\theequation}{2.\arabic{equation}}
\setcounter{equation}{0}

We shall consider the bosonic $\sigma$-model ($\mu,\nu=1, ..., D$)
\beq
&& Z = \int [\D x] \exp\big( -I(x)\big)\,,
\\
&& I = \frac{1}{4\pi\alpha'}\int d^2\sigma \sqrt{g} \Big( \frac{\alpha'}{\varepsilon^2}\varphi(x) + \partial_a x^\mu \partial^a x^\nu G_{\mu\nu}(x) +\alpha' R^{(2)}\phi(x)\Big)\,,
\eeq
defined on a compact closed two-dimensional surface. Here $G_{\mu\nu}$, $\phi$, and $\varphi$ are the 
bare fields that depend on the two-dimensional cutoff $\varepsilon$. The renormalized value of $\varphi$ will be chosen to be  zero. The theory is defined by the action $I$ and the measure $[\D x]$. Imposing the requirement of invariance under the  general coordinate transformations
$$
x^\mu \rightarrow x'{}^\mu\,, \qquad G_{\mu\nu} \rightarrow G_{\mu\nu}' = \frac{\partial x^\alpha }{\partial x'{}^\mu} \frac{\partial x^\beta}{\partial x'{}^\nu} G_{\alpha\beta}\,, 
$$
(i.e. that upon a transformation  of $x^\mu$ the ``coupling constants" $G_{\mu\nu}$ of the theory are also transformed), below we shall consider three ways of calculating the partition function (2.1) that are consistent with the requirement of this covariance.

\

\def \ov {\over}

{\bf{1}.} Let us first  choose  the measure $[\D x]$  to be trivial:
$$
[\D x]= \prod_\sigma d^\D x(\sigma)\,.
$$
To cancel the power divergences we make use of the bare tachyon field $\varphi(x)$ (with  the renormalized value of $\varphi$ set  to zero). We separate $x^\mu$ into a constant and a non-constant parts, 
 $x^\mu=y^\mu+\eta^\mu$,  inserting   ``one" into (2.1) (cf. \cite{12})
\beq
&& 1 = \int d^Dy \int \prod d^D\eta \,\delta^{(D)}(x(\sigma)-y-\eta)\ \delta^{(D)}(P^\mu[y,\eta])\ Q[y,\eta]\,,
\nonumber\\
&& Q = \det { \partial P^\mu[y-a,\eta+a]\ov \partial a^\nu}\Big|_{a=0}\,,
\eeq
where $P=0$ is a gauge condition and $Q$ is the ghost determinant. 
One possible  choice  is 
\be
P^\mu = \int d^2\sigma \sqrt{g}\ \eta^\mu\,,\qquad Q=V^D\,, \qquad V = \int d^2\sigma\ \sqrt{g} \,.
\ee
The condition $P=0$ implies that $\eta$ does not contain a zero mode of the Laplace operator (a constant). We substitute $x=y+\eta$ into the action and expand it in powers of $\eta$:
\beq
&& I = \frac{1}{4\pi\alpha'}\int d^2\sigma \sqrt{g} \Big[ \frac{\alpha'}{\varepsilon^2}\varphi + \partial_a \eta^\mu \partial^a \eta^\nu \big( G_{\mu\nu} +\partial_\lambda G_{\mu\nu} \eta^\lambda + \half \partial_\lambda \partial_\rho G_{\mu\nu}\eta^\lambda\eta^\rho+\ldots\big)
\nonumber\\
&& \hspace{5cm} +\alpha' R^{(2)} ( \phi + \half \partial_\mu\partial_\nu \phi \eta^\mu\eta^\nu+\ldots\big) \Big]\,.
\eeq
The leading (one-loop) contribution of  the integral over $\eta$ is 
\be
Z_0 = [\det{}' (G_{\mu\nu}\Delta)]^{-1/2} = \exp\big[ - \half N' \ln G - \half D \ln \det{}'\Delta\big]\,,
\ee
where $N'$ is the regularized number of nonzero eigenmodes of the Laplace operator,
 $G=\det G_{\mu\nu}$ and $D$ is the dimensionality of space-time. 

The number $N'$ can be expressed in terms of  the heat kernel  in a familiar way
\beq
&& N'= \int d^2\sigma\ \sqrt{g}  \,  K_\varepsilon -1 = \frac{V}{4\pi\varepsilon^2} + \frac{1}{6}\chi + O(\varepsilon^2)\,,
\\
&& K_\varepsilon = \sum_n f_n(\sigma)f_n(\sigma') \exp(-\lambda_n \varepsilon^2)\,,
\eeq
where $f_n(\sigma)$ and $\lambda_n$ are, respectively, the eigenfunctions and eigenvalues of the Laplace operator on the two-dimensional surface of Euler number $\chi= {1\ov 4 \pi} \int d^2 \sigma\sqrt g  R^{(2)}$.

Taking (2.7) into account, we obtain for (2.6)
\beq
&& Z_0 = \AA \exp\Big[\big(-\frac{V}{4\pi\varepsilon^2} - \frac{1}{6}\chi + 1 + O(\varepsilon^2)\big) \ln G\Big]\,,
\nonumber\\
&& \AA = \exp\big[-\half D \ln \det{}'\Delta\big]\,.
\eeq
The dependence of $Z$ on the dilaton field (to order $\alpha'^2$) is easily found from (2.5):
\be
Z = \int d^D y\ \ZZ_0\ e^{-\chi\phi}\big[ 1 - \alpha'\pi \chi \partial_\mu \partial_\nu \phi\ G^{\mu\nu}\ \DD(\sigma,\sigma) + O(\alpha'{}^2)\big]\,,
\ee
where ${\cal D}$ is the regularized Green function of the Laplace operator 
\be
\DD(\sigma,\sigma') = \sum_{ \lambda_n\ne 0} \frac{f_n(\sigma)f_n(\sigma')}{\lambda_n} \exp\big(-\lambda_n\varepsilon^2)\,.
\ee
For $\varepsilon\rightarrow 0$ it has the form \cite{13}
\be
\DD(\sigma,\sigma) = - \frac{1}{2\pi} \ln \varepsilon + O(1)\,.
\ee
To determine the dependence of $Z$ on the graviton field $G_{\mu\nu}$ it is necessary to consider the two possible one-particle-irreducible two-loop diagrams. Their contribution to $Z$ is found to be 
\beq
&& Z = \int d^Dy\ \ZZ_0 \ e^{-\chi\phi}\Big[ 1 + c_1 G^{\mu\nu}G^{\lambda\rho} \partial_\lambda \partial_\rho G_{\mu\nu} + c_2 G^{\mu\alpha} G^{\nu\beta}G^{\rho\lambda}  \partial_\rho G_{\mu\nu} \partial_\lambda G_{\alpha\beta} \qquad
\nonumber\\
&& \hspace{4cm} + c_3 G^{\mu\lambda} G^{\nu\beta}G^{\rho\alpha}  \partial_\rho G_{\mu\nu} \partial_\lambda G_{\alpha\beta}  + O(\alpha'{}^2)\Big]\,,
\nonumber\\
&& c_1 = - \half \pi\alpha' \DD(\sigma,\sigma)N'\,,
\nonumber\\
&& c_2 =  \half \pi\alpha' \int d^2\sigma \ d^2\sigma' \sqrt{g}\sqrt{g'} \  \DD(\sigma,\sigma')\  \partial_a \partial{_{b'}}\DD(\sigma,\sigma')\  \partial^a\partial^{b'}\  \DD(\sigma,\sigma') \,,
\nonumber\\
&& c_3 =  \pi\alpha' \int d^2\sigma \ d^2\sigma'\  \sqrt{g}\sqrt{g'}\ \partial^a\DD(\sigma,\sigma') \ \partial_a\partial_{b'} \DD(\sigma,\sigma') \ \partial^{b'} \DD(\sigma,\sigma')\,.
\eeq
We ensure the covariance of $Z_0$ by means of the  special choice of the bare fields
\be
\phi' = \phi + a \ln \sqrt {G(x)}\,, \qquad \varphi' = \varphi + b \ln \sqrt {G(x)}\,.
\ee
Substituting  $x=y+\eta$ and  expanding  $\ln\sqrt {G(x)}$ in powers of $\eta$  we  obtain the following correction to the action in (2.2) 
\beq
&& \Delta I = \frac{1}{4\pi\alpha'} \int d^2\sigma \sqrt{g}\  \alpha' \ \Big( \frac{b}{\varepsilon^2} + a R^{(2)} \Big) \Big( \ln \sqrt{G} + \frac{1}{4} G^{\mu\nu} \partial_\lambda \partial_\rho G_{\mu\nu} \eta^\lambda \eta^\rho\qquad
\nonumber\\
&& \hspace{6cm} -\,\, \frac{1}{4} G^{\mu\beta} G^{\nu\alpha}\partial_\rho G_{\mu\nu} \partial_\lambda G_{\alpha\beta}\eta^\rho\eta^\lambda + \ldots \Big)\,.
\eeq
The values of $a$ and $b$ are calculated from the condition that $Z_0$ has a 
required covariant form $(Z_0\sim\sqrt{G})$
\be
a = - \frac{1}{6}\,, \qquad b=-1\,.
\ee
We now find the correction to $Z$ from (2.15)  taking into account (2.16) and the final expression for $Z_0$ 
\beq
&& \Delta Z = \AA \int d^Dy\ \sqrt{G}\  e^{-\chi\phi} \Big[ 1 + \half\pi\alpha' \big(\frac{1}{6}\chi +\frac{V}{4\pi\varepsilon^2} \big) \DD(\sigma,\sigma)
\nonumber\\
&& \hspace{3cm} \times \,\, \Big( G^{\mu\nu} G^{\lambda\rho} \partial_\lambda \partial_\rho G_{\mu\nu} - G^{\mu\alpha} G^{\nu\beta} G^{\rho\lambda}  \partial_\rho G_{\mu\nu} \partial_\lambda G_{\alpha\beta} + O(\alpha'{}^2)\Big)\Big]\qquad
\eeq
As a result, from (2.9), (2.10), (2.13), and (2.17), we obtain 
\beq
&& Z = \AA \int d^Dy \sqrt{G} e^{-\chi\phi}\Big( 1 - \pi \alpha' \chi \DD(\sigma,\sigma) \partial_\mu\partial_\nu\phi G^{\mu\nu} +  \tilde{c_1}G^{\mu\nu} G^{\lambda\rho} \partial_\lambda\partial_\rho G_{\mu\nu}
\nonumber\\
&& \hspace{1cm} + \,\, \tilde{c}_2 G^{\mu\alpha} G^{\nu\beta}G^{\rho\lambda}  \partial_\rho G_{\mu\nu} \partial_\lambda G_{\alpha\beta} +  \tilde{c}_3 G^{\mu\lambda} G^{\nu\beta}G^{\rho\alpha}  \partial_\rho G_{\mu\nu} \partial_\lambda G_{\alpha\beta} + O(\alpha'{}^2)\Big)\ \,,
\nonumber\\
&& \tilde{c_1} = c_1 + \half \pi\alpha' (\frac{1}{6} \chi + \frac{V}{4\pi \varepsilon^2})\DD(\sigma,\sigma)\,,
\nonumber\\
&& \tilde{c_2} = c_2 - \half \pi\alpha' (\frac{1}{6} \chi + \frac{V}{4\pi \varepsilon^2})\DD(\sigma,\sigma)\,,\qquad \tilde{c_3} = c_3\,.
\eeq
After the $c_i$'s have been calculated using  (2.7) and (2.12), the power divergences cancel and the dependence of $Z$ on $\varepsilon$ takes the form 
\beq
&& Z = \AA \int d^Dy\, \sqrt{G}\, e^{-\chi\phi}\, \Big[ 1 +\half  \alpha' \chi(\ln \varepsilon + O(1))\partial_\mu\partial_\nu\phi \, G^{\mu\nu}
\nonumber\\
&& \hspace{3cm} - \,\, \frac{1}{4}\alpha' (\ln \varepsilon +O(1)) G^{\mu\nu} G^{\lambda\rho} \partial_\lambda\partial_\rho G_{\mu\nu}
\nonumber\\
&& \hspace{3cm} + \,\, \frac{1}{8}\alpha' (\ln \varepsilon +O(1)) G^{\mu\alpha} G^{\nu\beta}G^{\rho\lambda}  \partial_\rho G_{\mu\nu} \partial_\lambda G_{\alpha\beta}
\nonumber\\
&& \hspace{3cm} + \,\, \frac{1}{4}\alpha' (\ln \varepsilon +O(1)) G^{\mu\lambda} G^{\nu\beta}G^{\rho\alpha}  \partial_\rho G_{\mu\nu} \partial_\lambda G_{\alpha\beta} +
O(\alpha'{}^2)\Big]
\eeq
 Using in  (2.19) the expression for the target space scalar 
curvature $R$ in terms of $G_{\mu\nu}$  and integrating by parts   we observe that we can rewrite $Z$  in the manifestly covariant form 
\be
Z = \AA \int d^Dy\ \sqrt{G}\ e^{-\chi\phi}\ \Big[ 1 + \half \alpha'\big(\ln \varepsilon + O(1)\big)\big(R+ \chi \rd^2 \phi\big) + O(\alpha'{}^2)\Big]\,,
\ee
where $\rd_\mu$ in $\rd^2$  is the covariant derivative. 
\

{\bf{2}.}  Next, let us  consider the manifestly covariant 
 method of calculating $Z$ based on  the 
 expansion for the action and the measure in normal coordinates. 
 Let us   define the measure $[\D x]$   by the formal product 
\be
[\D x]= \prod_\sigma d^D x(\sigma)\ \sqrt{ G(x(\sigma))}\,.
\ee
To preserve the general covariant invariance in the regularized theory it is necessary to regularize the measure and the action in a consistent manner. We choose the regularized expression for the measure (2.21) in the form 
\beq
&& [\D x]= \prod_\sigma d^D x(\sigma)\  e^\MM
\\
&& \MM = \half \int d^2\sigma \sqrt{g} \ln G(x) K_\varepsilon(\sigma,\sigma)\,.
\eeq
Now let  us set $x^\mu = y^\mu  + \eta^\mu(y,\xi) $  where $\xi^\mu$ is the tangent vector to the geodesic joining the points $y^\mu$ and $y^\mu+\eta^\mu$ 
\be
\eta^\mu= \xi^\mu - \half \Gamma_{\alpha\beta}^\mu \xi^\alpha\xi^\beta - \frac{1}{6} \big(\partial_\gamma \Gamma_{\alpha\beta}^\mu - 2 \Gamma_{\gamma\alpha}^\lambda \Gamma_{\lambda\beta}^\mu\big) \xi^\alpha\xi^\beta\xi^\gamma+\ldots
\ee
\def \rD {{\rm D}}
The expansions of the action and measure in powers of $\xi$ have the form \cite{14}
\begin{align}
& I = \frac{1}{4\pi\alpha'} \int d^2\sigma \sqrt{g}\Big[ \partial_a \xi^\mu \partial^a \xi^\nu \Big( G_{\mu\nu} +\frac{1}{3} R_{\mu\lambda\rho\nu}\xi^\lambda \xi^\rho + \frac{2}{45} R_{\lambda\mu\rho}{}^\gamma R_{\alpha\nu\beta\gamma} \xi^\lambda \xi^\rho \xi^\alpha \xi^\beta + O(\xi^5)  \Big)\qquad  %
\nonumber\\
&  \hspace{5cm}
+ \alpha' R^{(2)}\Big( \phi + \rD_\mu \phi \, \xi^\mu + \half \rD_\mu \rD_\nu \phi\,  \xi^\mu \xi^\nu + O(\xi^3)\Big)\Big]\,,
\end{align}
\be
\MM = \half \int d^2\sigma \sqrt{g} \ K_\varepsilon(\sigma,\sigma) \Big( \ln G - \frac{1}{3} R_{\mu\nu}\xi^\mu\xi^\nu + O(\xi^3)\Big)\,.
\ee
Since the kinetic term is invariant under a  constant shift  $\xi\rightarrow \xi+a$
and $\xi$  may  contain a constant part under the condition (2.4), 
it is desirable to fix the symmetry $y\rightarrow y-a,\  \eta\rightarrow\eta+a$ by means of another gauge condition \cite{12}
\be
P^\mu = \int d^2\sigma \sqrt{g}\ \xi^\mu\,.
\ee
In this case the ghost determinant in (2.3)  is 
\be
Q = \det\Big(\int d^2\sigma \sqrt{g}\ \lambda^\mu{}_\nu\Big)\,,\ \ \  \qquad \lambda^\mu{}_\nu = \frac{\partial \xi^\mu(y,\eta)}{\partial \eta^\nu } - \frac{\partial \xi^\nu(y,\eta)}{\partial \eta^\mu }\,.
\ee
Its covariant expression takes the form 
\be
Q = V^D \exp\Big(-\frac{1}{3V} \int d^2\sigma \sqrt{g} \ R_{\mu\nu}\xi^\mu\xi^\nu + O(\xi^3)\Big)\,.
\ee
 To determine the measure in the $y$ integral, i.e.  $\int d^Dy\sqrt{G}$, 
  it is necessary to take into account not only the one-loop contribution (2.6) but also (2.26). Using 
  that the regularized  number of eigenvalues is 
\be
N = \int d^2\sigma \sqrt{g}\ K_\varepsilon(\sigma,\sigma)\,,
\ee
and also (2.7), we arrive at the  expression for the  covariant 
\ measure $\sqrt{G}$ in the integral over $y$. In fact, $\sqrt{G}$ is the contribution of the only  (constant) 
 zero mode of the Laplace operator on the compact surface. The partition function $Z$ then  takes the form 
\be
Z = \int d^Dy \sqrt{G}\ e^{-\chi \phi}\  F(R,\rD R,\rD\phi)\,.
\ee
It is not difficult to calculate the first terms of the expansion of $F$ in powers of $\alpha'$. From (2.25), (2.26), and (2.29), we obtain 
\be
Z =  \ZZ_0 \int d^Dy \sqrt{G}\,  e^{-\chi \phi}\Big[ 1 + \alpha'(a_1+a_2+a_3)R + \alpha' b_1 \rD^2\phi + O(\alpha'{}^2)\Big]\,.
\ee
The coefficients $a_1$ and $b_1$ correspond to contributions from the action (2.25), $a_2$ arises from the measure (2.26), and $a_3$ from the ghost determinant (2.29). The expressions for these coefficients in terms of the Green functions (2.11) have the following appearance
\beq
&& a_1 = \frac{\pi}{3} \int d^2\sigma\sqrt{g}\ \partial_a \partial'^{a} \DD(\sigma,\sigma')|_{\sigma=\sigma'} \DD(\sigma,\sigma) - \partial_a \DD(\sigma,\sigma')|_{\sigma=\sigma'}  \partial'^{a} \DD(\sigma,\sigma')|_{\sigma=\sigma'}\,,\qquad
\nonumber\\
&& a_2 = -\frac{\pi}{3} \int d^2\sigma\sqrt{g} K_\varepsilon(\sigma,\sigma)
\DD(\sigma,\sigma)\,,
\\
&& a_3 = -\frac{2\pi}{3V} \int d^2\sigma\sqrt{g} \DD(\sigma,\sigma)\,,\qquad
b_1 = -\frac{1}{4} \int d^2\sigma\sqrt{g} R^{(2)}\DD(\sigma,\sigma)\,.
\nonumber
\eeq
Explicit calculations give 
\beq
&& a_1 = -\frac{1}{6} N'\ln\varepsilon + \bar{a}_1\,, \qquad a_2 = \frac{1}{6} N\ln\varepsilon + \bar{a}_2\,,\qquad a_3 = \frac{1}{3} \ln\varepsilon + \bar{a}_3\,,
\nonumber\\
&& b_1 = \half \chi \ln\varepsilon + \bar{b}_1\,, \qquad a_0 = a_1 + a_2+a_3 = \half \ln\varepsilon + \bar{a}_0\,,
\eeq
where the $\bar a_i$ and $\bar b_i$ are finite constants. It is easy to see that the power infinities cancel, and the resulting expression for $Z$ in (2.32) coincides with (2.20). 

\

{\bf 3}.  Let us  now  consider one more 
method of calculating $Z$, which is explicitly covariant and turns out to be simpler in practice. 
Here  we  define the measure $[\D x]$ as follows
\be
[\D x] = J\, d^Dy\,  [\D\xi]\,,
\ee
where the factor $J$ is fixed from the normalization condition
\beq
&& \int [\D\delta x] \ e^{-||\delta x||^2} = \int d^D\delta y\int  [\D\delta\xi ] \,  J \, e^{-||\delta x||^2}=1\,,
\nonumber\\
&& ||\delta x||^2 = \frac{1}{4\pi\alpha'} \int d^2\sigma \sqrt{g}\,  \delta x^\mu \delta x^\nu\,  G_{\mu\nu}\,.
\eeq
The expression for $||\delta x||^2$ expanded in normal coordinates has the form 
\beq
&& ||\delta x||^2 = \frac{1}{4\pi \alpha'} \int d^2\sigma \sqrt{g} \Big( G_{\mu\nu} + \frac{1}{3} R_{\mu\lambda_1\rho_1\nu} \xi^{\lambda_1}\xi^{\rho_1}
\nonumber\\
&& \hspace{1.5cm} - \frac{2}{45} R_{\mu\lambda_1\rho_1}{}^{\gamma_1} R_{\alpha_1 \nu\beta_1\gamma_1} \xi^{\lambda_1} \xi^{\rho_1} \xi^{\alpha_1} \xi^{\beta_1} + \ldots\Big) \Big(\delta y^\mu + \delta \xi^\mu + \frac{1}{3} R^\mu{}_{\lambda_2\rho_2\kappa} \xi^{\lambda_2}\xi^{\rho_2} \delta y^\kappa
\nonumber\\
&& \hspace{1.5cm} - \frac{1}{45} R^\mu{}_{\lambda_2\rho_2\gamma_2} R^{\gamma_2}{}_{\alpha_2 \beta_2\kappa} \xi^{\lambda_2} \xi^{\rho_2} \xi^{\alpha_2} \xi^{\beta_2}\delta y^\kappa + \ldots \Big)
\Big(\delta y^\nu + \delta \xi^\nu + \frac{1}{3} R^\nu{}_{\lambda_3\rho_3\sigma} \xi^{\lambda_3}\xi^{\rho_3} \delta y^\sigma
\nonumber\\
&& \hspace{1.5cm} - \frac{1}{45} R^\nu{}_{\lambda_3\rho_3\gamma_3} R^{\gamma_3}{}_{\alpha_3 \beta_3\sigma} \xi^{\lambda_3} \xi^{\rho_3} \xi^{\alpha_3} \xi^{\beta_3} \delta y^\sigma + \ldots \Big)\ . 
\nonumber
\eeq
Integrating successively over $\delta y$ and $\delta\xi$, we find $J$. Taking into account the expression (2.35) for the action, we have 
\begin{align}
& Z  =  \ZZ_0  \int d^Dy \sqrt{G} e^{-\chi \phi} \Big< \exp\Big[ \int d^2\sigma \sqrt{g} \Big( \frac{\pi\alpha'}{3} R_{\mu\lambda\nu\rho} \partial_a \xi^\mu \partial^a\xi^\nu \xi^\lambda\xi^\rho - \frac{\pi\alpha'}{3}  K_\varepsilon'(\sigma,\sigma) R_{\mu\nu} \xi^\mu \xi^\nu\qquad
\nonumber\\
& \hspace{1cm} -\,\, \frac{\pi\alpha'}{V} R_{\mu\nu}\xi^\mu\xi^\nu - \frac{4\pi^2\alpha'{}^2}{45} R_{\lambda\mu\rho}{}^\gamma R_{\alpha\nu\beta\gamma} \partial_a\xi^\mu \partial^a \xi^\nu \xi^\lambda\xi^\rho\xi^\alpha\xi^\beta
\nonumber\\
& \hspace{1cm}  +\,\, \pi^2 \alpha'{}^2\big( \frac{4}{45} K_\varepsilon'(\sigma,\sigma) + \frac{2}{3V} \big) R_{\mu\lambda\rho}{}^\gamma R{}^\mu{}_{\alpha\beta\gamma} \xi^\lambda \xi^\rho\xi^\alpha\xi^\beta + \ldots\Big)
\nonumber\\
& \hspace{1cm}  - \,\, \int d^2\sigma d^2\sigma' \sqrt{g}\sqrt{g'} \pi^2\alpha'{}^2\Big( \frac{1}{9} K_\varepsilon'{}^2(\sigma,\sigma') + \frac{2K_\varepsilon'(\sigma,\sigma')}{9V} + \frac{1}{V^2}\Big)
\\
& \hspace{5.5cm}  \times R_{\mu\rho\lambda\nu} R^\mu{}_{\alpha\beta}{}^\nu \xi^\lambda(\sigma)\xi^\rho(\sigma) \xi^\alpha(\sigma')\xi^\beta(\sigma') + O(\alpha'{}^3)\Big]\Big>\,.\nonumber
\end{align}
We have redefined $y\rightarrow(2\pi\alpha')^{1/2}y$ and $\xi\rightarrow (2\pi\alpha')^{1/2}\xi$, set  $\phi$ to be constant for simplicity, and took the one-loop contribution into account. Starting from (2.37), we easily find the  expression for the
order  $\alpha'$ terms in  $Z$. It  is given by the first three terms in the exponent in  (2.37). 
The contribution of the second term cancels that of the first one so that the coefficient of $R$ turns out to be proportional to ${\cal D}(\sigma,\sigma)$,  so that as in (2.20) we get 
\be
Z =  \ZZ_0   \int d^Dy \sqrt{G} \, e^{-\chi \phi}\,  \Big( 1 + \half  \alpha' (\ln\varepsilon  + \hbox{const} )R + O(\alpha'{}^2) \Big)\,.
\ee
The divergent parts of the coefficients of the $R^2$ and $R_{\mu\nu}^2$ terms are calculated in a similar way. 
One gets for the  $R^2$ term 
\be
Z = \ZZ_0 \int d^Dy \sqrt{G} e^{-\chi \phi} \Big( 1 + \ldots + \half \pi^2 \alpha'{}^2 \DD^2(\sigma,\sigma) R^2 + \ldots\Big)\,.
\ee
The divergent  contribution to 
 the coefficient of  the $R_{\mu\nu}^2$ term  comes  effectively only from 
  the vertex $-{\pi^2\alpha'^2\ov V} R^2\xi\xi\xi\xi$, i.e.
\be
Z = \ZZ_0 \int d^Dy \sqrt{G} e^{-\chi \phi} \Big( 1 + \ldots   - \pi^2 \alpha'{}^2 \DD^2(\sigma,\sigma) R^{\mu\nu} R_{\mu\nu} + \ldots\Big)\,.
\ee

\

The   methods  of computing  $Z$ considered above admit a natural generalization to the case of 
2d  surfaces with boundaries (with   free  open string or Neumann boundary conditions).
 Then the  Green function $\DD$  is replaced by the  Neumann function.
 There are  
 new (linear) power divergencies  
  which can be canceled by a redefinition of the values of the boundary analogs of the tachyon and dilaton  couplings.
   The  $\sigma$-model  action in this case has the form 
\be
I = \frac{1}{4\pi\alpha'}\int d^2\sigma\sqrt{g}\Big(\frac{\alpha'\varphi}{\varepsilon^2} + \partial_a x^\mu \partial^a x^\nu G_{\mu\nu} + \alpha' R^{(2)} \phi\Big) +{1\ov 2 \pi}  \int ds \Big(\frac{\varphi'}{\varepsilon} + {K} \phi'\Big)\,,
\ee
where $K$ is the extrinsic curvature. 
It is necessary to set $\phi=\phi'$ to ensure that  the  constant part of the dilaton couples to the   Euler characteristic.

It should be emphasized that  the above  calculation of  $Z$ 
was done for surfaces of any genus. 
However, we did not integrate over the moduli space of the Riemann surfaces and, therefore, the logarithmic divergences found are only the ordinary local ones. 

The expression  $Z$ is renormalizable with respect to these local infinities on a surface of  an arbitrary genus
($\psi^i=(G, \phi)$)
\be
\frac{d Z}{d\ln\varepsilon} = \frac{\partial Z}{\partial\ln\varepsilon} -
\beta^i \frac{\partial Z}{\partial\psi^i} = 0\,,
\ee
where  $\beta^i=- {d\ov d \ln \varepsilon} \psi^i$  are the  local $\beta$-functions of the $\sigma$-model (cf. (2.20))  
\be
\beta^G_{\mu\nu} = \alpha' R_{\mu\nu} + O(\alpha'{}^2)\,, \qquad \beta^\phi =  \frac{1}{6}D - \half \alpha' \rd^2 \phi +  O(\alpha'{}^2)\, .
\ee

Assuming that $Z$ is renormalizable  also at the next order 
and using the known expressions for the $\alpha'^2$ terms in the $\beta$-functions (2.43)  \cite{12,15}, we find the
following expression for the  logarithmically divergent term  in $Z$ to order $\alpha'^2$ 
\be
Z = \lambda\int d^Dy\,  \sqrt{G} \, e^{-\chi \phi}\,  \Big[ 1 +\half  \ln \varepsilon \Big(  \alpha' R +
 \frac{1}{8} (4-\chi) \, \alpha'{}^2R_{\mu\alpha\beta\nu} R^{\mu\alpha\beta\nu}   \Big)+\ldots \Big]\,.
\ee
We shall  also confirm   the coefficient  of the $R_{\mu\alpha\beta\nu} R^{\mu\alpha\beta\nu}$ term directly 
in the case of the torus ($\chi=0$)  in the next section. 

Let us  note also that the $(\ln \varepsilon)^2$ 
coefficients of $R^2$ and $R_{\mu\nu}^2$ that we found in (2.39) and (2.40)  are consistent with the  renormalizability  of $Z$.

\section{Partition function  on specific 2-surfaces: sphere, disk and torus}
\renewcommand{\theequation}{3.\arabic{equation}}
\setcounter{equation}{0}

Let us now  consider the calculation of $Z$ for some  simplest surfaces: the sphere, disk, and torus. In these cases the coefficients of the leading terms in the $\alpha'$ expansion of $Z$ can be found explicitly. 

{\bf 1}. Let us start with  the 2-sphere. In spherical coordinates the eigenfunctions and eigenvalues of the Laplace operator have the form 
\be
f_{n,m} = Y_{n,m} (\theta,\phi)\,, \qquad \lambda_{n,m} = n(n+1)\,,
\ee
where the $Y_{n,m}$ are the orthonormal  spherical functions. The regularized expression for the
 Green's  function has the form \be
\DD(\sigma,\sigma') = \sum_{n\ne 0} \sum_{m=-n}^n \frac{1}{n(n+1)} e^{-n(n+1)\varepsilon^2}\, Y^*_{n,m}(\theta,\varphi)Y_{n,m}(\theta',\varphi')\,.
\ee
At coincident points, it becomes 
\be
\DD(\sigma,\sigma) = \frac{1}{4\pi} \sum_{n\ne 0} \frac{2n+1}{n(n+1)} e^{-n(n+1)\varepsilon^2} \ . 
%
\ee
The leading terms in expansion in $\varepsilon\to 0$ are easily calculated using the Euler-Maclaurin resummation formula 
\be
\DD(\sigma,\sigma) = - \frac{1}{2\pi}\ln \varepsilon + \frac{\gamma-1}{4\pi} + \frac{\varepsilon^2}{6\pi} + O(\varepsilon^4)\,,
\ee
where $\gamma$ is the Euler constant. We note that the $\ln\varepsilon$ and $\varepsilon^2$ terms can be calculated from (2.7) using integration over $\varepsilon$. Taking (3.4) into account, we can write (2.20) as (here $\chi=2$)
\be
Z = \ZZ_0 \int d^D y \sqrt{G} e^{-2\phi}\Big[ 1 + \alpha'\big( R + 2\rd^2\phi)\Big(\half  {\ln\varepsilon} + a + O(\varepsilon^2)\Big) + O(\alpha'{}^2)\Big]\,,
\ee
where $a=\frac{1}{4} (\gamma-1)$ is a scheme-dependent constant.

It is easy to see that $Z$ is renormalizable, i.e. making the replacement $G_{\mu\nu}=G^{(R)}_{\mu\nu}-\ln\varepsilon\,\beta_{\mu\nu}^{G}$ and $\phi=\phi^{(R)}-\ln\varepsilon\,\beta^{\phi}$ (cf. (2.43)) 
 we get rid of the logarithmic divergences   and thus find 
\be
Z = \ZZ_0 \int d^D y \sqrt{G} e^{-2\phi}\Big[ 1 + a \alpha'\big( R + 2\rd^2\phi) + O(\alpha'{}^2)\Big]\,.
\ee
Note that this  expression is not the same as  the closed string effective action 
obtained using  the S-matrix method.
 The reason is that the generating functional for the string tree-level  S-matrix  is given by 
   $\Omega^{-1} Z$, i.e  $Z$ divided by the volume of the group $SL(2,\CC)$ of M\"obius transformations. The presence of a logarithmic singularity in the regularized volume of $SL(2,\CC)$ suggest that one can think 
   of $\partial\ov \partial\ln\varepsilon$ as a possible realization of the operation of division by $\Omega$ in the case of closed strings \cite{4}. Indeed, as follows from (3.5), 
\be
\frac{\partial Z}{\partial \ln\varepsilon} = \half \alpha' \ZZ_0   \int d^D y\,  \sqrt{G} \, e^{-2\phi}\, \Big[ R + 2\rd^2\phi  + O(\alpha')\Big]\,,
\ee
which  agrees  with the  effective action found from the tree-level closed  string S-matrix. 

\

{\bf 2}. The calculation of $Z$ for  disk topology  (with a metric of half-sphere) 
 almost analogous to the case of the  sphere. 
A new  feature is that in view of the presence of the boundary, we impose  the  Neumann boundary condition
at the boundary of half-sphere
\be
\partial_\theta x^\mu \big|_{\theta =\frac{\pi}{2}} = 0\,.
\ee
The expansion of the fluctuation field $\eta$ in eigenfunctions of the Laplace operator on the disk has the form 
\be
\eta(\theta, \phi)  = \sum_{n,m} a_{n,m} Y_{n,m}\,, \qquad n+m = 2k\,, \qquad n\ne 0\,.
\ee
and the expression for the regularized Neumann function at coincident points is (cf. (3.3))
\be
\DD(\sigma,\sigma) = \sum_{n=1}^\infty \frac{1}{2\pi n } \ e^{-n(n+1)\varepsilon^2}\,.
\ee
Using the Euler-Maclaurin formula, we obtain (cf. (3.4))
\be
\DD(\sigma,\sigma) = -\frac{1}{2\pi} \ln \varepsilon  + \frac{\gamma}{4\pi} + \frac{1}{4}\varepsilon - \frac{5}{12\pi} \varepsilon^2+ O(\varepsilon^3)\,.
\ee
The expression for $Z$ is the same as in (2.18), (2.20) with $\DD(\sigma, \sigma) $   given by  (3.11). The power divergences $\varepsilon^{-2}$ and $\varepsilon^{-1}$ in $Z$ on the disk are canceled by renormalizing the tachyon fields $\varphi$ and $\varphi'$, respectively (see (2.41)).

\

\def \F  {{\rm F}}

{\bf 3}. In the case of the 2-torus we shall depart from the scheme used above, which was based on
the heat kernel
 regularization. 
 This is due to the technical difficulties of calculating the sums with the  spectral 
  $e^{-\lambda_n\varepsilon^2}$ regularization.\footnote{Note that in \cite{16} the authors used  a regularization 
   based on a cutoff on the upper limits of the sums over  eigenmodes of the Laplace operator on the torus.}
   We shall consider the $\tau$-parametrization, in which the torus 
   is represented as a $(\tau, 1)$ parallelogram on the complex $z$-plane. The string $\sigma$-model 
   partition function on the torus 
   has the form \cite{17}
\beq
&&Z =\int_\F \frac{d^2\tau}{4\pi\tau_2^2} \frac{e^{4\pi\tau_2}}{(2\pi\tau_2)^{12}} |f(e^{2\pi\irm \tau})|^{-48} \int \D x\,  e^{-I}\,, \\ && \int \D x\,  e^{-I_0}=1\,, \qquad I = I_0 + I_{int}\,, \qquad  f(e^{2\pi\irm \tau}) = \prod_{n=1}^\infty ( 1 - e^{2\pi\irm n \tau})\ , \nonumber 
\eeq
where the fundamental region $\F$ is specified by the conditions
$$
- \half < \tau_1 \leq \half, \qquad  |\tau| > 1 \,,\qquad \tau= \tau_1 + i \tau_2 \ .
$$
We shall  consider only  the dependence of $Z$ on the metric $G_{\mu\nu}$. 
By studying the dependence of $Z$ on $G_{|mu\nu}=\delta_{\mu\nu}+h_{\mu\nu}$ using 
the  expansion in powers of $h_{\mu\nu}$, we will  then restore the coefficients of the $R$, $R^2$, $R_{\mu\nu}^2$ and $R_{\mu\alpha\beta\nu}^2$ terms (assuming  that the scheme used for the regularization and renormalization preserves the covariance of $Z$).

Since the metric on the parallelogram is flat, it is possible to use the following regularization prescription 
$$
\DD(z,z) = - \frac{1}{2\pi} \ln \varepsilon\,, \qquad \delta^{(2)}(z,z)=0\,, 
$$
corresponding to  discarding of  power divergences. 
This prescription ensures the covariance of $Z$ without  need  for a nontrivial measure factor. 
The Green function on the torus has the form \cite{17}
\be
\DD(z,z') = -\frac{1}{4\pi}\ln \frac{|\theta (z,z')|^2 }{|\theta'(0)|^2} + \frac{1}{2\tau_2}\big[ {\rm Im}(z-z')\big]^2\,,
\ee
where $\theta(z,z')$ is the theta function $\vartheta_{11}(z,z')$ \cite{18}.

We redefine $x\rightarrow(2\pi\alpha')^{1/2}x$ and expand the $\sigma$-model   action  $I$  in (2.2) 
in powers of $\eta=x-y$. Then 
\beq
&& Z = \langle \ZZ \rangle \ , \qquad \ZZ= 
\int \D\eta \exp\Big[- \half \int d^2 \sigma \sqrt{g} \partial_a \eta^\mu \partial^a \eta^\nu\Big(\delta_{\mu\nu}  +
h_{\mu\nu} +  (2\pi\alpha')^{1/2} \partial_\lambda h_{\mu\nu} \eta^\lambda %
\nonumber\\
&& \hspace{3cm} + \pi\alpha' \partial_\rho \partial_\lambda h_{\mu\nu} \eta^\rho \eta^\lambda + \ldots \Big) \Big] \,, \qquad \langle\ldots\rangle  = \int d^D y \int  [d\tau]\,.
\eeq
The coefficient of $R$ is found  from the  $h_{\mu\nu}\square h_{\mu\nu}$ 
term 
($R={1\ov 4}h_{\mu\nu}\square h_{\mu\nu}+\dots$). As a result, 
\beq
&& \ZZ \sim 1 + 2\pi\alpha' c_0 R + O(\alpha'{}^2)\,,
\nonumber\\
&& c_0 = 4 \int d^2z d^2z' \big( \partial_z\partial_{z'} \DD \partial_{\zb}\DD\partial_{\zb'}\DD + \partial_z\partial_{\zb'} \DD \partial_{\zb}\DD\partial_{z'}\DD
\nonumber\\
&& \hspace{2cm} + \partial_{\zb}\partial_{\zb'} \DD \partial_z\DD\partial_{z'} \DD  + \partial_{\zb}\partial_{z'} \DD \partial_z \DD\partial_{\zb'}\DD\big)\,.
\eeq
Integrating by parts and using the regularization indicated above, we obtain 
$$
c_0 = \frac{1}{4\pi} \ln \varepsilon + O(1)\,, \qquad \ZZ \sim  1 + \half \alpha'\big(\ln \varepsilon +O(1)\big)R + O(\alpha'{}^2)\,.
$$
To calculate the coefficients of the $R^2$, $R_{\mu\nu}^2$, and $R_{\mu\alpha\beta\nu}^2$ terms we note that 
\be
aR^2 + bR_{\mu\nu}^2 + c R_{\mu\alpha\beta\nu}^2 = (a + \half b + c) \partial_\mu \partial_\nu h^{\mu\nu} \partial_\alpha\partial_\beta h^{\alpha\beta} + (a+\frac{1}{4}b) \partial^2 h_\mu{}^\mu\partial^2 h_\alpha{}^\alpha\, + ...
\ee
On the other hand, the coefficient $a$ is in fact known (it is related to the coefficient of the $R$ term), since $R$ and $R^2$ effectively arise from the expansion of the exponential $e^R$. Thus 
\be
a = \frac{1}{8} \ln^2\varepsilon   + O(\ln\varepsilon)\,.
\ee
We note that  like the finite part of the $c_0$ in (3.15), the coefficients  of $\ln \varepsilon$  terms in 
 $a$ and $b$
 are not unique, i.e. depend on a 
  regularization scheme.\footnote{Note  that  the ambiguity of the $\ln\varepsilon$ terms in $a$ and $b$ does not  affect the coefficient $c$ as  $a+\half  b\sim O(1)$.} Finding the coefficient $\lambda_1=a + \half b+c$ and $\lambda_2=a+{1\ov 4} b$ in (3.16) and using (3.17), we can  calculate $b$ and $c$. Expanding (3.14) to order $\partial^4h$, we obtain
\beq
&& \lambda_1 = 32 \int d^2z d^2z' \partial_z \DD \partial_{\zb} \DD \partial_{z'}\DD  \partial_{\zb'} \DD\,,
\nonumber\\
&& \lambda_2 = 4 \int d^2z d^2z' \partial_z \partial_{\zb}\DD\big|_{z=z'} \partial_{z'}\partial_{\zb'} \DD\big|_{z=z'} \big(\DD^2(z,z) + \DD^2(z,z'))\,.
\eeq
From this it  follows that 
\beq
&& \lambda_1 = \frac{1}{4} \ln\varepsilon + O(1)\,, \qquad \lambda_2 = \frac{1}{16} \ln^2\varepsilon + O(\ln\varepsilon)\,,
\nonumber\\
&& b = - \frac{1}{4} \ln^2\varepsilon + O(\ln\varepsilon)\,, \qquad c =  \frac{1}{4} \ln\varepsilon + O(1)\,.
\eeq
Thus, the expression for $Z$ has the form
\beq
&& Z = \ZZ_1 \int [d\tau] \int d^Dy \sqrt{G} \Big( 1 + \half \alpha' \ln \varepsilon R
\nonumber\\
&& \hspace{2cm} +\,\,  \frac{1}{8} \alpha'{}^2 \ln^2 \varepsilon R^2 -  \frac{1}{4} \alpha'{}^2 \ln^2 \varepsilon R_{\mu\nu}^2 +  \frac{1}{4} \alpha'{}^2 \ln \varepsilon R_{\mu\alpha\beta\nu}^2+ \ldots \Big)\,.
\eeq
The coefficient of $R_{\mu\alpha\beta\nu}^2$ is consistent with the renormalizability of $Z$ (cf. (2.44)  with $\chi=0$ and (2.42)), as it is easily seen from the well known expression  \cite{12} for the two-loop $\beta^G$-function 
$$
0 = \frac{dZ}{d\ln\varepsilon} = \frac{\partial Z}{\partial \ln\varepsilon} -
\beta^G_{\mu\nu} \frac{\partial Z}{\partial G_{\mu\nu}} \,, \qquad \beta_{\mu\nu}{}^G = \alpha' R_{\mu\nu} + \half \alpha'{}^2 R_{\mu\alpha\beta\gamma} R_\nu{}^{\alpha\beta\gamma} + O(\alpha'{}^3)\,.
$$
Note that,  in fact,  we have effectively 
 calculated the local $\beta^G$-function of the $\sigma$-model on a torus.
  It coincides with that on a sphere, as expected. 
  The direct calculation of $\beta^G$ on a torus was also performed in \cite{16}. Compared to \cite{16} where 
    cumbersome expressions arose and cutoff  regularization of the sums was applied, 
    our calculation using   $Z$ is rather simple.
     We should stress that the possibility of deriving $\beta^G$ from $Z$ is a distinctive feature of the torus geometry:
      there is an $R^2$ term in the dilaton $\beta^{\phi}$ as well, but
      for the torus  the $e^{-\chi\phi}$ factor is trivial as  $\chi=0$. 

The above method of calculating $Z$  illustrated on the example of the torus   which is 
based on the use of the trivial measure for $x^\mu$, 
an expansion in $h_{\mu\nu}=G_{\mu\nu}-\delta_{\mu\nu}$, and a special prescription for subtracting 
power divergences that ensures the covariance of $Z$, is  closest in spirit 
 to the usual method of calculating string scattering amplitudes as correlators of vertex operators. 
 
 This approach can be generalized to 
  surfaces of higher  genus (where, to ensure invariance it is necessary to discard $\delta^{(2)}(z,z)$ altogether, i.e.
   to discard the $\varepsilon^{-2}$ divergence and the finite part ${1\over 6}\chi$ term in (2.7)). 
   Integrating by parts in (3.18), one can prove that the prescription $\delta^{(2)}(z,z)=0$ is sufficient to
    verify the universality of the coefficients of the $\ln^2\varepsilon$ terms in $a$ and $b$ in (3.16).
     Note that though the value of the coefficient of $\ln\varepsilon$ in $\lambda_2$ in the general case depends on a choice of regularization, the value of $c(4-\chi)\ln\varepsilon$ is the same in all regularizations that preserve the covariance (for example, in dimensional and in $\delta^{(2)}(z,z)=0$ regularizations).

\section{Partition function of the $N=1$ supersymmetric $\sigma$-model }
\renewcommand{\theequation}{4.\arabic{equation}}
\setcounter{equation}{0}

Let us  generalize the results of Sec.2 to the case of the supersymmetric 2d $\sigma$-model
related to fermionic (NSR)  string in curved background. 
The important difference from the bosonic case is the automatic cancellation of power UV  divergences.

The action of a fermionic string in flat space is given by (see, e.g., \cite{19})
\be
I = \frac{1}{2\pi\alpha'} \int d^4z\,  E\,  D_- \hat{x}^\mu D_+\hat{x}_\mu\,,
\ee
where $ d^4z E=d^2\sigma d\theta d\bar\theta\, \text{sdet} E^A_M$, $\hat x^\mu$ is a scalar superfield, $D_{-}$ and $D_{+}$ are superderivatives, and $(\sigma_1,\sigma_2, \theta,\bar\theta)$ are the coordinates on the supersurface. 

For the action of the  corresponding supersymmetric $\sigma$-model we have  
\be
I = \frac{1}{2\pi\alpha'} \int d^4z\, E \, D_- \hat{x}^\mu D_+\hat{x}^\nu\,  G_{\mu\nu}(\hat{x})
+ \frac{\irm }{2\pi} \int d^4z\,  E\,  R_{+-}\,  \phi(\hat{x}) \,.
\ee
Here $R_{+-}$ are the  components of the two-dimensional curvature tensor, and $G_{\mu\nu}$ and $\phi$ are the graviton and dilaton fields. Note that the Euler characteristic can be written also  as 
\be
\chi =  \frac{\irm }{2\pi} \int d^4z\,  E\,  R_{+-} \,.
\ee
The component expansion of $\hat x^\mu$ is 
\be
\hat{x}^\mu = x^\mu + \theta^r \psi_r^\mu + \irm \theta \bar\theta F^\mu\,.
\ee
We shall use the antiperiodic boundary conditions for the field $\psi$ 
\be
\psi(\varphi+2\pi) = - \psi(\varphi)\,.
\ee
Here $\varphi$ is the polar angle in the complex plane or angle of a cylinder.
 In this case the Dirac operator does not have zero modes (but  the scalar Laplace operator has). On surfaces of higher genera this choice of boundary conditions corresponds to an even spin structure for $\psi$ 
\be
\psi(z+a_i) = - \psi(z)\,,\qquad \psi(z+b_i) = - \psi(z)\,,
\ee
where $i=1,\dots, g$, and $a_i$ and $b_i$ are the basis cycles on the Riemann surface. 

We shall use the supersymmetric generalization of the heat kernel method used in Sec.2. The expressions (2.7) and (2.8) become 
\beq
\hat{N}' = \int d^4 z\,  E\, \hat{K}_\varepsilon - 1= \frac{1}{2} \chi - 1 + O(\varepsilon^2)\,,
\\
\hat{K}_\varepsilon = K_\varepsilon(\sigma,\sigma) \delta^2(\theta, \theta) = \frac{\irm }{4\pi} R_{+-} + O(\varepsilon^2)\,.
\eeq
Note that $-1$ in (4.7) corresponds to the bosonic zero mode $y^\mu=const$. As already mentioned, 
in  contrast to the bosonic case, here the $\varepsilon^{-2}$ divergence is absent  which is a manifestation of the  two-dimensional supersymmetry  which also forbids the standard  tachyon term in the $\sigma$-model  action  (cf. (2.2)).

To calculate $Z$ we separate in $\hat x$ the zero mode, $\hat x=y+\hat\eta$, $\D\hat x=d^Dy\, \D\hat\eta$. The terms in the action (4.2) that contribute in the one-loop approximation have the form 
\be
I = \frac{1}{2\pi\alpha'} \int d^4z\, E \, D_- \hat{\eta}^\mu D_+\hat{\eta}^\nu G_{\mu\nu}(y)
+ \frac{\irm }{2\pi} \int d^4z\, E\, R_{+-} \phi(y)\,.
\ee
Analogously to (2.9), we obtain 
\beq
&& \hat{Z}_0 = \hat{\ZZ_0}\exp\Big[\big(1-\frac{1}{2}\chi + O(\varepsilon^2)\big)\ln G\Big]\,,
\nonumber\\
&& \hat{\ZZ_0} = \exp (-\half D \ln  \det{}' \hat{\Delta})\,.
\eeq
As in the bosonic case, the factor $(\sqrt{G})^\chi$ can be absorbed into a redefinition of the dilaton field 
\be
\phi' = \phi + a\ln \sqrt{G(\hat{x})}\,.
\ee
The value of $a$ is fixed by the condition that $\hat Z$   should be   covariant. 
As a result, $a=-\half$. 

To find $\hat Z$ in the two-loop approximation, we choose the integration measure as (cf. (2.35))
\be
D \hat{x} = J\, d^Dy\,  \D\hat{\xi}\,,
\ee
where $J$ is determined from the normalization condition
\be
\int \D\delta \hat{x} e^{-||\delta\hat{x}||^2} =1\,, \qquad ||\delta\hat{x}||^2 = \frac{}{} \int d^4z\, E\, \delta \hat{x} \delta\hat{x}^\nu\, G_{\mu\nu}\,.
\ee
Performing the calculation analogous to the one  in the bosonic case and using the 
normal coordinates $\hat\xi$, we get 
\beq
&& \hat{Z} = \hat{\ZZ_0} \int d^D y\,  \sqrt{G}\,  e^{-\chi\phi}\, \Big< \exp\Big(\int d^4 z\,  E\  \frac{1}{3} \pi\alpha' R_{\mu\alpha\nu\beta} D^\gamma \hat{\xi}^\mu D_\gamma \hat{\xi}^\nu \hat{\xi}^\alpha \hat{\xi}^\beta
\nonumber\\
&&\qquad \qquad\qquad   - \pi\alpha' \Big(\frac{1}{3} \hat{K}_\varepsilon'(z,z) + \frac{1}{V}\Big)R_{\alpha \beta} \hat{\xi}^\alpha \hat{\xi}^\beta+ O(\alpha'{}^2)\Big)\Big>\,,
\eeq
where $\langle ...\rangle$ is computed with the free gaussian  action for  the normal coordinate fields $\hat{\xi}^\alpha $.
As in the bosonic case, we have redefined $\hat\xi\rightarrow(2\pi\alpha')^{1/2}\hat\xi$, taken the one-loop contribution into account, and  have chosen $\phi$=const. As a consequence, 
\beq
&& \hat{Z} =   \hat{\ZZ_0}    \int d^D y\,  \sqrt{G}\,  e^{-\chi\phi}\, \Big[ 1 - \pi\alpha' \hat{\DD}(z,z) R + O(\alpha'{}^2)\Big]\,.
\eeq
Using the regularized expression for $\hat{\DD}(z,z)$ (see, e.g.,  \cite{19})
$$
\hat{\DD}(z,z) = - \frac{1}{2\pi} \ln \varepsilon + O(1)\,,
$$
(4.14) becomes
\be
\hat{Z} =   \hat{\ZZ_0}  \int d^D y\,  \sqrt{G}\,  e^{-\chi\phi}\, 
\Big[1 + \half\alpha' \big( \ln \varepsilon + O(1)\big) R + O(\alpha'{}^2)\Big]\,,
\ee
which (at this leading  oder in $\alpha'$) coincides with the  bosonic string expression  in (2.20).

For the case of the sphere with a nontrivial dilaton field  we get
\be
\hat{Z} =  \hat{\ZZ_0} \int d^D y \, \sqrt{G} \, e^{-2\phi}\,
 \Big[1 + \half\alpha' \big( \ln \varepsilon + O(1)\big) \Big(R +2G^{\mu\nu} \rd_\mu \rd_\nu \phi\Big ) + O(\alpha'{}^2)\Big]\,.
\ee
Applying the $\partial\ov \partial\ln\varepsilon$ prescription for ``dividing" over the volume of the  super-M\"obius 
group  we obtain 
\be
\frac{\partial \hat{Z}}{\partial \ln\varepsilon} =  \half \alpha'  \hat{\ZZ_0} \int d^D y \, \sqrt{G}\,  e^{-2\phi}\,\Big (R +2 \rd ^2 
\phi  + O(\alpha')\Big)\,,
\ee
that agrees with the expression  for the  superstring effective action (same as bosonic action to this order in (3.7)) 
 found using  the S-matrix approach.

\newpage


\end{document}